\documentclass[12pt,preprint]{aastex}
\usepackage{url}
\usepackage{rotating}
 \usepackage{graphics,graphicx}
 \usepackage{psfig}

\newcommand{\Msun}{\mbox{$M_{\odot}$}}

\newcommand{\Swift}{\mbox{{\it Swift}}}

\begin{document}
\title{The Supernovae Analysis Application (SNAP)}
\author{Amanda J. Bayless\altaffilmark{1},  Chris L. Fryer\altaffilmark{3, 4, 5}, Brandon Wiggins\altaffilmark{3,6}, Wesley Even\altaffilmark{3},  Ryan Wollaeger\altaffilmark{3}, Janie de la Rosa\altaffilmark{2}, Peter W. A. Roming\altaffilmark{1, 2},   Lucy Frey\altaffilmark{3},   Patrick A. Young\altaffilmark{7}, Rob Thorpe\altaffilmark{1}, Luke Powell\altaffilmark{1}, Rachel Landers\altaffilmark{1}, Heather D. Persson\altaffilmark{1}, Rebecca Hay\altaffilmark{1}}

\altaffiltext{1}{Southwest Research Institute, Department of Space Science, 6220 Culebra Rd, San Antonio, TX 78238, USA}
\altaffiltext{2}{University of Texas at San Antonio, San Antonio, TX 78249, USA}
\altaffiltext{3}{Los Alamos National Laboratory, Los Alamos, NM 87545, USA}
\altaffiltext{4}{Physics Department, University of Arizona, Tucson, AZ 85721, USA}
\altaffiltext{5}{Physics and Astronomy Department, University of New Mexico, Albuquerque, NM 87131, USA}
\altaffiltext{6}{Southern Utah University, 351 W University Blvd, Cedar City, UT 84720, USA}
\altaffiltext{7}{School of Earth and Space Exploration, Arizona State University,  411 N Central Ave, Phoenix, AZ 85004, USA}

\begin{abstract}

The SuperNovae Analysis aPplication (SNAP) is a new tool for the analysis of SN observations and validation of SN models. SNAP consists of an open source relational database with (a) observational light curve, (b) theoretical light curve, and (c) correlation table sets, statistical comparison software, and a web interface available to the community.  The theoretical models are intended to span a gridded range of parameter space. The goal is to have users to upload new SN models or new SN observations and run the comparison software to determine correlations via the web site. There are looming problems on the horizon that SNAP begins to solve.  Namely, large surveys will discover thousands of SNe annually. Frequently, the parameter space of a new SN event is unbounded. SNAP will be a resource to constrain parameters and determine if an event needs follow-up without spending resources to create new light curve models from scratch.  Secondly, there is not a rapidly available, systematic way to determine degeneracies between parameters or even what physics is needed to model a realistic SNe.  The correlations made within the SNAP system begin to solve these problems.    

\end{abstract}


\section{Introduction}

One of the most crucial processes for shaping the composition of the universe is the death of massive stars manifest as supernovae (SNe). Understanding this process is important as it has significant impacts of our understanding of cosmology, chemical enrichment, galaxy evolution star formation rate, stellar evolution, compact object remnants, circumstellar medium and dust formation.  SNe also serve as laboratories of high-energy, high-density regimes, which cannot be probed effectively by experiments on earth. 

SNe are classified by the evolution of their emission and spectral features (for reviews, see \citealt{fil05,fil97}) with Types Ib, Ic, and II being believed to arise from the collapse of the core of a massive star and subsequent stellar explosion.  Observational differences between core-collapse events likely arise from either variations in the progenitor star or the circumstellar environment or possibly characteristics of the explosion mechanism itself  (e.g. \citealt{arn96,heg03}). Multiple code bases (e.g. \citealt{bli93,des05,kasen2006,wol13,fre13}) are solely dedicated to modeling the evolution of SN luminosity or SN light curves and recently efforts have begun to do detailed numerical comparisons of these codes \citep{koz16}. Previously, interesting SNe have been modeled individually (e.g. \citealt{des15,cha16}) and such studies have provided constraints on e.g. mass loss and stellar evolution (\citealt{bay15,fre13})or insights into the explosion mechanism itself \citep{gre14}.  Calculating emission can be an involved process, in some cases requiring $\sim 10,000$ CPU hours per light curve , with large suites of simulations dedicated to studying a single explosion \citep{fre13}.

Upcoming instruments, including LSST, will open the era of large SN surveys in which $>100,000$ SNe of all types, core-collapse and Type Ia, will be discovered annually \citep{val12}. To process this flood of SN data, it becomes imperative to have a means of quickly and robustly characterizing events.  This wealth of data will include previously rare or exotic explosions and robustly identifying these events for extended follow-up is essential to capitalizing. On-the-fly, detailed numeric modeling for every event is impractical, yet very high-fidelity models which include shock radiation hydrodynamics or detailed non-LTE radiative transfer may still be needed to identify SNe exhibiting a particular process of interest. These SN data will also provide a possible means of validating SN light curve codes over very large parameter spaces, provided each observed SNe can be studied inexpensively. Packages for fitting and characterizing SNe exist in the literature (e.g. \citealt{bur11}), but such applications make \textit{a priori} assumptions about light curve shape or nature of the power source.  There is also an existing archive for SN observations \citep{gui16}, but it does not contain models or any method of correlation.  In this paper we present the development of the SuperNovae Analysis aPplication (SNAP); an on-line resource including a database of high fidelity SN light curve models for characterizing and studying observed SNe. 

\Swift\ provides UV light curves of SNe, which are a critical probing ground for SN light curve models \citep{pri14}. Lack of UV early time observations cause an underestimation in the bolometric light curve by $\sim$20\% and up to 50\%. Fortunately, the \Swift\ UV/Optical Telescope (UVOT; \citealt{rom00,rom04,rom05}) has over 50 core-collapse SNe (CCSNe) within its archive \citep{pri14} and form the backbone of the current observational database.  At the time of this publication, we are in the process of adding additional \Swift\ SNe and parsing through ground-based observations of the \Swift\ observed SNe located in the \citet{gui16} archive. The model portion of the database is presently populated with models from the Los Alamos National Laboratory (LANL) and TYCHO \citep{you05}. 

Ultimately, we would like a unified understanding of how CCSN sub-types are related. This work represents a substantial contribution to this end. In \citet{bay15} we show that the transition between the classical hydrogen dominated SNe and hydrogen poor SNe can be modeled by altering the mass of hydrogen in the outer shell of the star prior to explosion. This former study, however, did not fully explore the rich variety of CCSN events. This study brings together for the first time a large array of high-fidelity SN models and a robust means of correlating SN observables with computer simulations. This allows for exploration of a wider range CCSN events on an unprecedented scale.  In this paper, we focus on CCSNe, but we also acknowledge the need for correlating observables for Type Ia SNe. We are beginning to ass Type Ia observations to the database and will have LANL Ia models available in an updated version of SNAP. We hope SNAP will be a community tool and invite others to add models (CCSNe and Ia).

This paper is structured as follows. In section 2, we describe the SN database, including the archived observations and numerical light curve models. In section 3, we detail our method of determining matches and correlations between models and observations. The web tool and interface is described in section 4 and we conclude in section 5.

\section{The Supernovae Databases}
\begin{figure*}
\center 
\includegraphics[scale=0.55,angle=0]{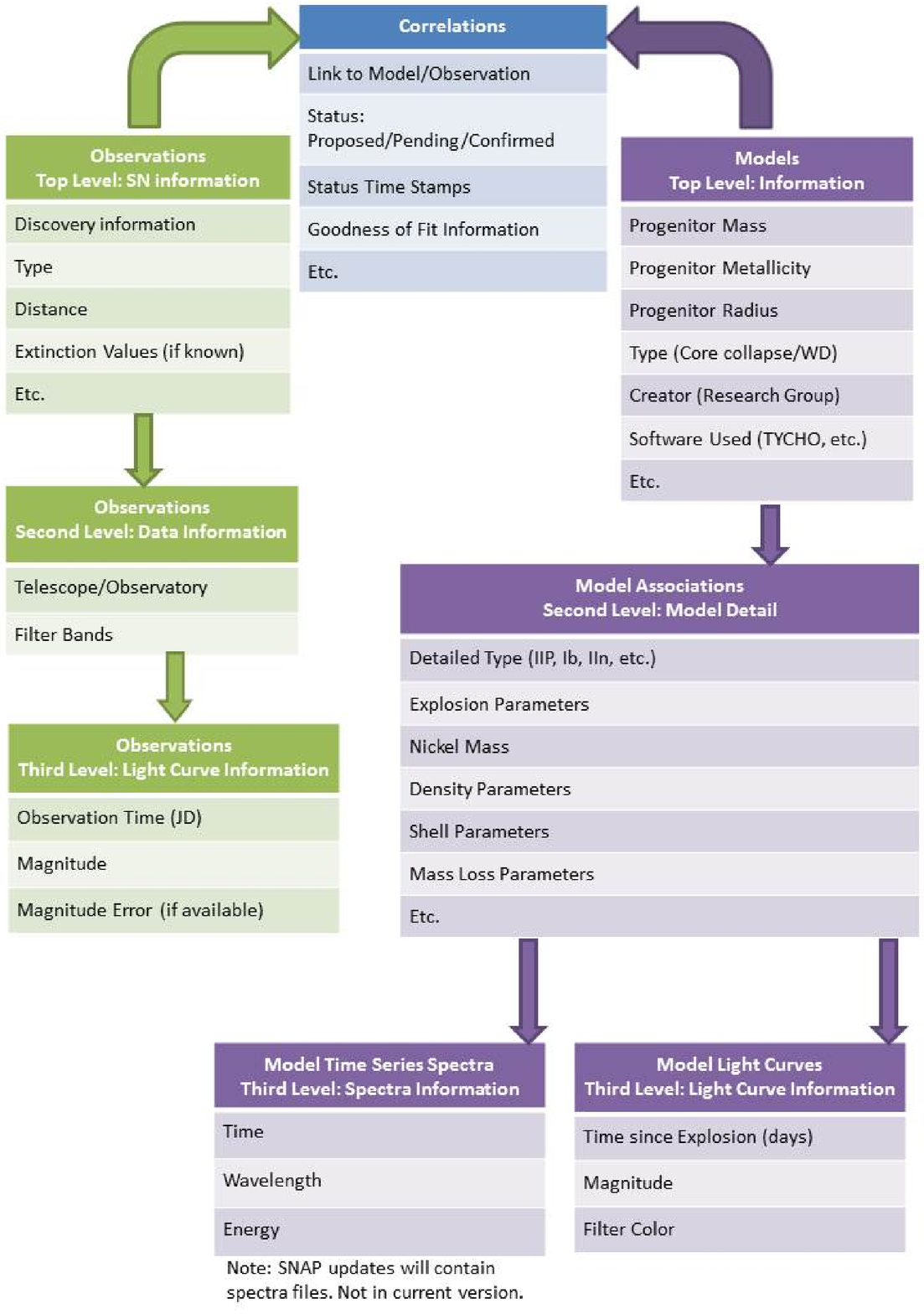}
\caption{A simplified SNAP database structure.  There are three primary data sets: 1) Observations, 2) Models, and 3) Correlations. The Observations are defined by a single SN event, which is further branched into one or more telescope/filter band combinations that observed the event and then branched into the light curve data. The models are uniquely defined by a progenitor star mass, metallicity, radius, and the code that created the model. The Model Associations are changes to that one progenitor star's explosion, density, mass loss, etc. This creates synthetic light curves and (in future versions of SNAP) time series spectra.  The correlations are saved observation-model comparisons that were deemed scientifically interesting. }
\label{snapchart}
\end{figure*}

The SNAP software consists of four main components: a) a relational database of models, observations, and correlations, b) a correlation software to compare models and observations, c) a web application and d) software functions to export observations and models and import correlations between observations and models.  The database component is the most important component of the application, providing a fully-relational, constraint-verified database structure for observations, models and correlations.  Figure \ref{snapchart} provides a high-level outline of the database structure.  The three main tables (Observations, Models, and Correlations) listed support SNAP science functions, but do not include system-level tables for managing users, recording audit records, etc. The Observation Table contains the observational data from the archives as light curve data sets. The Model Table contains the synthetic time series spectra and light curves from the LANL modeling code.  The Correlation Table is where the comparisons between the observations and models are stored.  The analysis will be in determining which data-model associations are scientifically interesting. 

The SNAP database is implemented in a MySQL relational database that contains twenty custom-designed tables that easily scale and support hundreds of thousands of observations, models, and correlations.  The web application functions come from the SwRI template web application, which provides pre-built web application capabilities such as user management, login authentication, database access, and many other functions.  This template web application has been used for dozens of missions including the Juno Science Operations Center, Cassini Ground Data Systems, and New Horizons-Ralph. The web application uses Java as the source software language, and runs on the open source Apache web server and Tomcat servlet container.  The I/O functions for observations, models and correlations are also being developed in Java, and currently run as stand-alone programs in the first version of SNAP, and will be fully integrated into SNAP in future versions.

\subsection{The Archived Light Curve Observations}
SNAP version 1.0 contains a subset of observations from the \Swift\ Observatory \citep{geh04}. We are currently updating SNAP to include more \Swift\ observations and light curves from other telescopes, both space and ground based, many of which are archived in \citet{gui16}.  We note that the difference between SNAP and the SN Open Source Catalog is that SNAP also includes models and a way to relate models and observations, and is not a repository for observations only. 

The \Swift\ Observatory includes the UVOT, which contains six filters ranging from the UV to optical \citep{poo08, bre11}. The \Swift\ archive is public, but recently the well observed UVOT SN light curves were published in a compilation paper \citep{pri14}.  This work demonstrated that the UV emission dominates the bolometric light curve up to 50 days post-explosion.  Therefore, including the UV data will be imperative to creating accurate models for SNAP.    There are generally multiple observations of the same SN event with which to correlate models to data and all of the available observations for an event are being compiled within SNAP.  Future SN light curves can immediately be input into SNAP and run with the fitting program against available light curve models to produce an instant estimation regarding the nature of the progenitor star and the explosion itself, and a realistic assessment of the error in that estimation.  As more SN models are added to the SNAP database the estimation in parameter space will improve. The benefit is that new SNe can be immediately compared to an existing model set, giving an estimate of parameter values, allowing for a determination of which new SNe should have follow-up observations. 

\subsection{The Light Curve Models}
SNAP 1.0 has been optimized for LANL SN models, described below.  However, the models are defined by several key parameters (e.g. progenitor mass, metallicity, explosion energy, etc.) that are fairly universal to all models.  Thus, SNAP has the ability to host most model systems. In the current version of SNAP, a light curve can be uploaded or the model data can be in the form of time series spectra.  Code is available to convert the spectra into \Swift\ band light curves for correlations with observations. Future versions of SNAP will include other standard band passes.  We note that the database stores only the resulting light curves/spectra and star/explosion parameters.  We do not store source codes, only the modeling outputs. We also recognize that even with the same parameters, different models can produce different results.  This feature of SNAP allows theoretical groups to compare light curves from same sets of parameters. The database has parameters giving the name of various codes used to make the model and include ``Stellar Evolution Software", ``Explosion Software",``Explosion Evolution Software", and ``Synthetic Spectra Software".  Each entry in the database also has an open notes field to include any other information.  More information regarding the LANL numerical model is a \citet{fre13} and https://ccsweb.lanl.gov/astro/transients/data/.

We currently have two modeling methods used in SNAP, both from LANL.  The first set of codes are described in detail in \citet{fre13} and summarized here. First, the initial parameters are chosen then we create a progenitor star using a stellar evolution code (TYCHO; \citealt{you05}).  The TYCHO progenitor stars also offer a new mixing algorithm (very different than the standard mixing length theory) based on solutions of the hydrodymanic equations of convection \citep{mea07,arn09,arn11}.  TYCHO also includes the options of large mass loss, e.g a Wolf-Rayet star \citep{lam02} or binary interaction, which will be needed for the SN Type Ib/c analysis. This new algorithm produces different cores that contain very little helium \citep{fre13b}.  These cores can drastically change the light curve predictions, particularly in the stripped-core SNe.  
The explosion and nuclear burning is modeled in a 1D Lagrangian code developed by \citet{her94}. This code includes three-flavor neutrino transport using a flux-limited diffusion calculation and a coupled set of equations of state to model the wide range of densities in the collapse phase (see \citealt{her94,fry99} for details). 

Evolution of SNe is done with the code Radiation Adaptive Grid Eulerian (RAGE; \citealt{git08}). The model from the 1D Lagrangian is mapped into RAGE where the material and radiation are simulated as they push out through the outer layers of the star and into the interstellar medium. RAGE is a multi-dimensional Eulerian radiation-hydrodynamics code with adaptive mesh refinement that can be used to model the physics in a wide range of SN sub-types.   One of the issues in modeling SNe lies in understanding the role of mass loss on the SN light curve, especially if the mass loss is episodic.  Interaction of the SN shock with this mass loss will play an important role in modeling SN light curves.  RAGE is ideally suited for such situations.  RAGE is also a two-temperature model with separate temperatures for the radiation and the matter.  As the simulation evolves, the grid mesh is increased to accommodate the ejecta expansion.  

 The hydrodynamic models are then post-processed with a separate code, SPECTRUM, to produce synthetic spectra.  SPECTRUM calculates luminosity using opacity data derived from the LANL OPLIB database \citep{mag95}, which contains data for a grid of 14,900 frequencies. The 1D RAGE output is mapped onto a 2D grid of radial and angular bins.  The luminosity of a grid element for a line of sight is determined by the blackbody temperature and the absorption, wavelength-dependent opacity, which is then attenuated by the optical depth.  The SPECTRUM code also accounts for the Doppler shifting of photons as they travel outward. Thus, the absorption opacity is determined from Doppler-shifted wavelengths.    Due to the lack of readily available models, many observers derive their first estimates of SN progenitor properties using a simple blackbody fit to the observed spectra (cf. \citealt{rom12}), assuming a single temperature and radius for the emitting surface.  \citet{fre13}  clearly illustrates the inaccuracy of this assumption.  Variations in temperature, density and composition with radius result in much more complex spectra and light curves than can be calculated using blackbody approximations. Once the spectrum is computed for several given simulation times, a synthetic light curve is computed for a given wavelength bin by applying a filter transmission curve to the spectrum.

\begin{figure*}
\center 
 \includegraphics[scale=0.45]{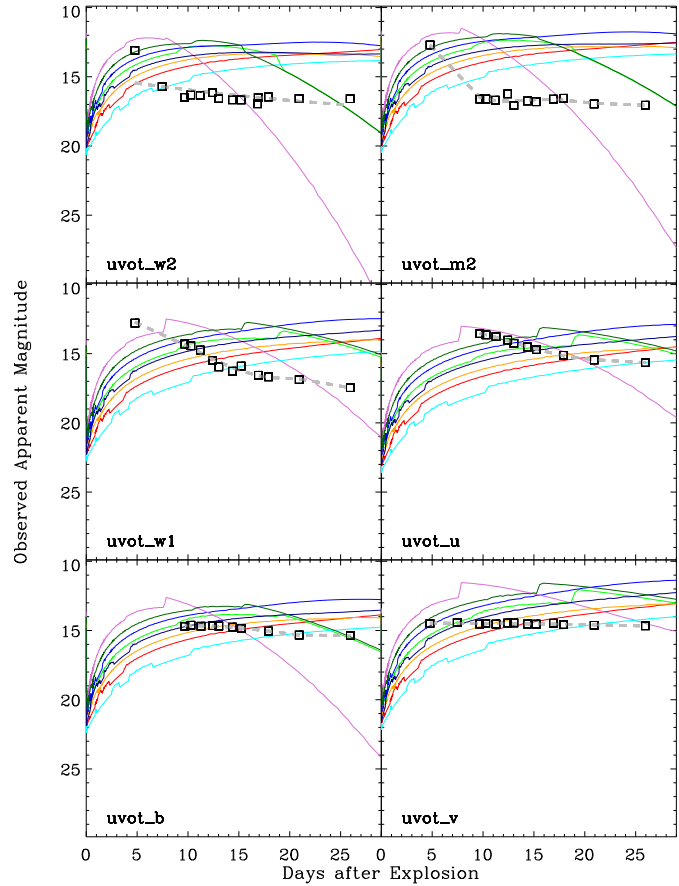}
\caption{Light curve models using the code from \citet{del16}. The open squares are SN 2008ax, a type IIb, from the optical {\it Swift}-UVOT band passes. The gray dashed line is the Gaussian process to the observation.  The colored lines are a set of models described in the text.}  
\label{correl}
\end{figure*}

The second set of codes use more simplified physics assuming a homologous outflow that models the transport in the gray diffusion approximation \citep{del16}.  While, this model is more simplistic, it is used because of its computational speed and, thus, the ability to create more light curves rapidly.  This also is useful to set parameter limitations before running the more elaborate code above.  To calculate bands under this approximation, we assume the emission of each zone is a blackbody and calculate the escape fraction of each zone near the photosphere to calculate a broad spectrum.  Because we do not include line features (more important at short wavelengths), this method will overestimate the emission in the UV and soft X-ray bands of \Swift\ (see Figure \ref{correl}).  Despite its weaknesses, this method is quick enough to calculate large grids of stars, varying the explosion energy, nickel yield, stellar mass and radius.  For this study, we varied only the energy and explosion mass, keeping the nickel yield fixed at 0.15 \Msun\ and the stellar radius at explosion of ~1.6d13 cm.

At the time of this publication, we are also working on alternative modeling codes that uses either the smooth particle hydrodynamics code SNSPH~\citep{fryer2006}
or the Monte Carlo semi-relativistic thermal radiative transfer
code SuperNu~\citep{wol13,wol14}.
SuperNu employs a hybrid of Implicit Monte Carlo 
\citep[IMC,][]{fleck1971,wollaber2016} for transport
and Discrete Diffusion Monte Carlo 
\citep[DDMC,][]{densmore2007,densmore2012,abd12} to accelerate
transport in optically thick regions of space and photon wavelength.
For SuperNu, the expansion is assumed to be homologous ~\citep[see, for
instance,][]{kasen2006}.
The mapping is performed when the ejecta is homologous (when
the particles are in a state of ballistic expansion).
We plan to incorporate multidimensional light curves from SuperNu in SNAP. In SNAP, these multi-dimensional light curves could allow
observers and theorists to examine which viewing angles of a
model best match an observation.

\section{Determining Best Correlations} 

The correlations are done by either comparing one model to the observations in the database or one observation (archived or new) to the suite of models in the database. In either case, the basic correlation is done the same way. A simple least-squares or similar fitting to determine the goodness-of-fit is not necessarily the best method for determining {\it how} a model fits or does not fit.  Knowing where the model does not fit data gives insight into producing better models.  For example, a model may have the same time of peak brightness and fall off slope from peak as an observation, but be too bright overall, missing the observational data points.  On the other hand, a straight line may pass through some data points, but this model would not be representative of the underlying physics of the SN.  A simple least-squares analysis would identify the straight line as the better fit even if the other model is a better representation.  A good light curve fit for a transient stems from modeling physical processes relevant in the SN emission. 

To determine a "goodness-of-physics" fit we follow the following steps.  First, we use either a Gaussian process, or if there are few data points a polynomial, to each observation in each filter.  We use the Gaussian process as it easy to incorporate observational noise, yields a variance at each fitting point which could be used in a reduced $\chi^2$,  does not have to conform to a particular functional form, and has a statistical "likely function" that can be maximized for optimal fitting parameters. We then account for distance and extinction and convert the model flux into the apparent magnitudes (currently for \Swift\ bands).  We use published values of the distance to the SNe and values for E(B-V) line-of-sight of the host galaxy (if available) from the literature on the particular SN event. The line-of-sight Milky Way extinction comes from Sloan Digital Sky Survey spectra of stars \citep{sch11} and the host galaxy extinction is from literature of individual events (see \citealt{pri14} and references therein). In the database itself, each event includes a reference field listing relevant publications. In this paper we use the extinction formula of \citet{car89} with a default value of $R_v = 3.1$, but $R_v$ is an adjustable parameter and the extinction law of \citet{cal00} is also a current option.  Figure \ref{correl}  shows \Swift\ observations with a set of simple physics LANL models described above.   This model has an ejected mass of 25 \Msun\ and explosion energy of 0.5 foe. 

 The code calculates a least-squares fit ($\chi^2$) for each filter and the sum of all filters by comparing the model woth Gaussian process (or polynomial) fit.  An added feature to the correlation code is the ability to shift the models in both magnitude and time interactively. There can be ambiguity in the explosion time\footnote{We note that in our database, time = 0 is defined as the shock breakout time.} and in the distance of the SN or if one wants to determine how to better fit a new model from an existing model. This ability to shift the models in time/magnitude space allows for some "$\chi$-by-eye" fitting rather than relying solely on automatic fits.  For each interactive shifting, the resulting $\chi^2$ values are output to a comma delimited file.  

\begin{figure*}[t]
\includegraphics[scale=0.45]{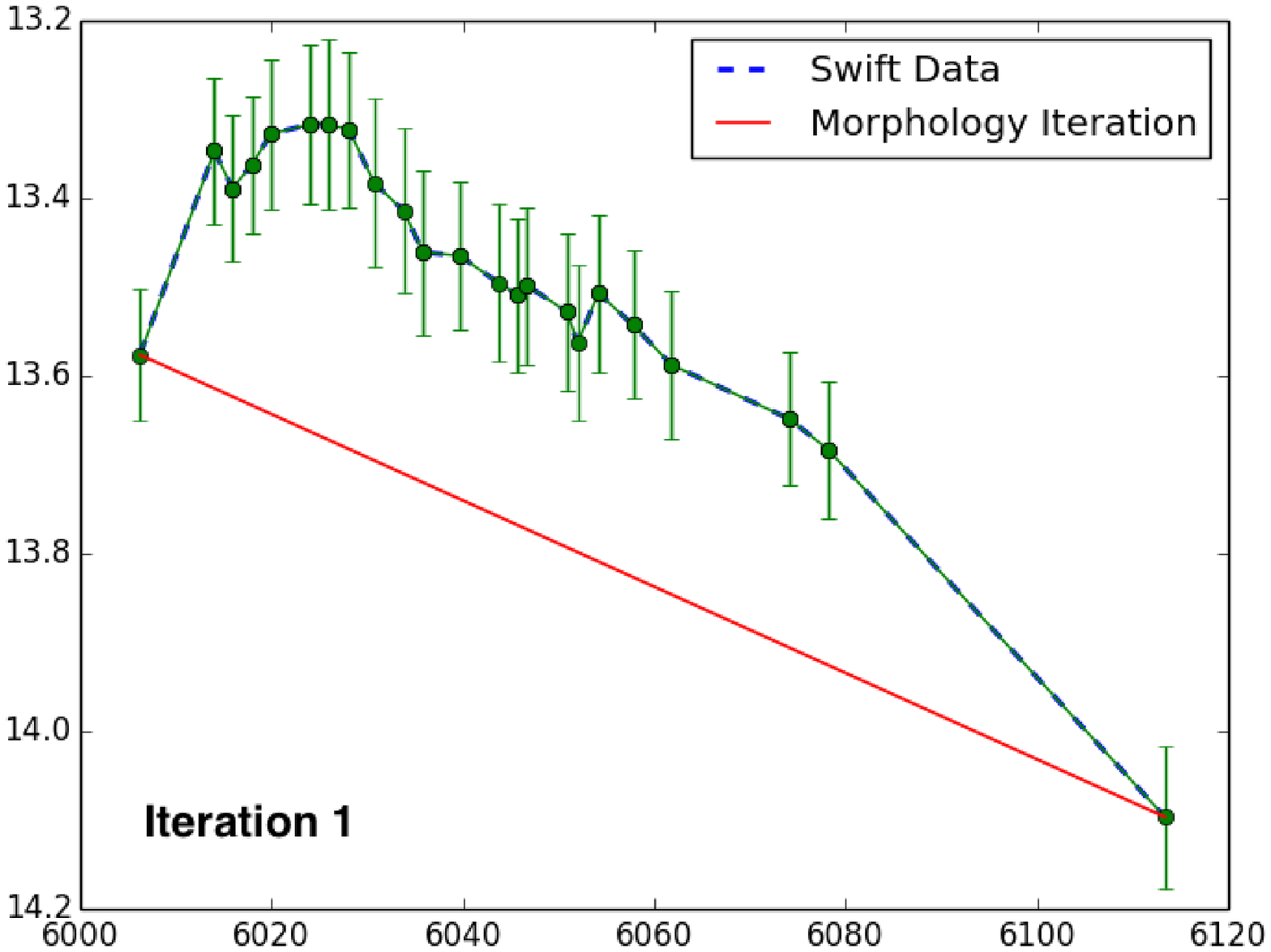}
\includegraphics[scale=0.45]{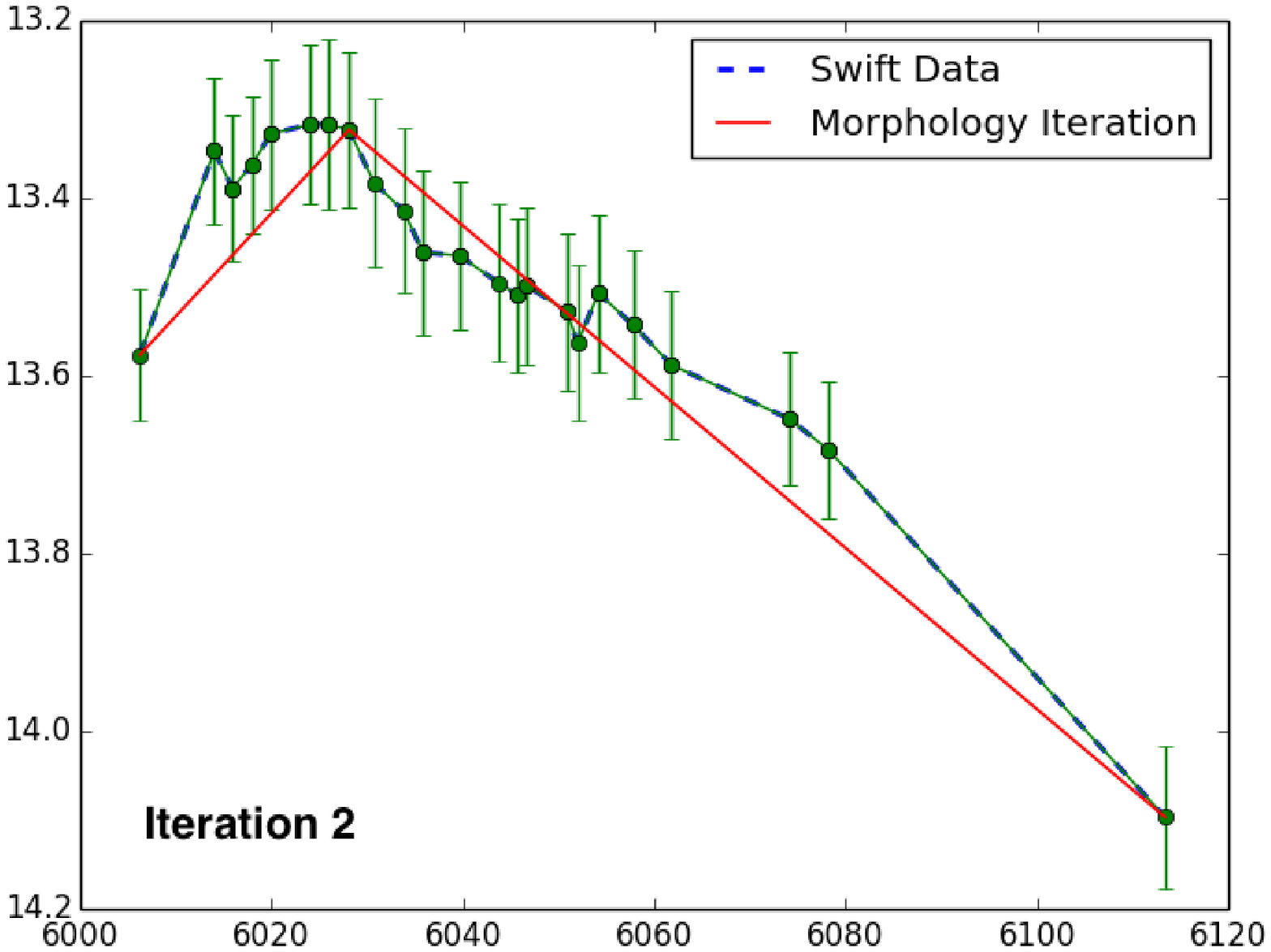}
\includegraphics[scale=0.45]{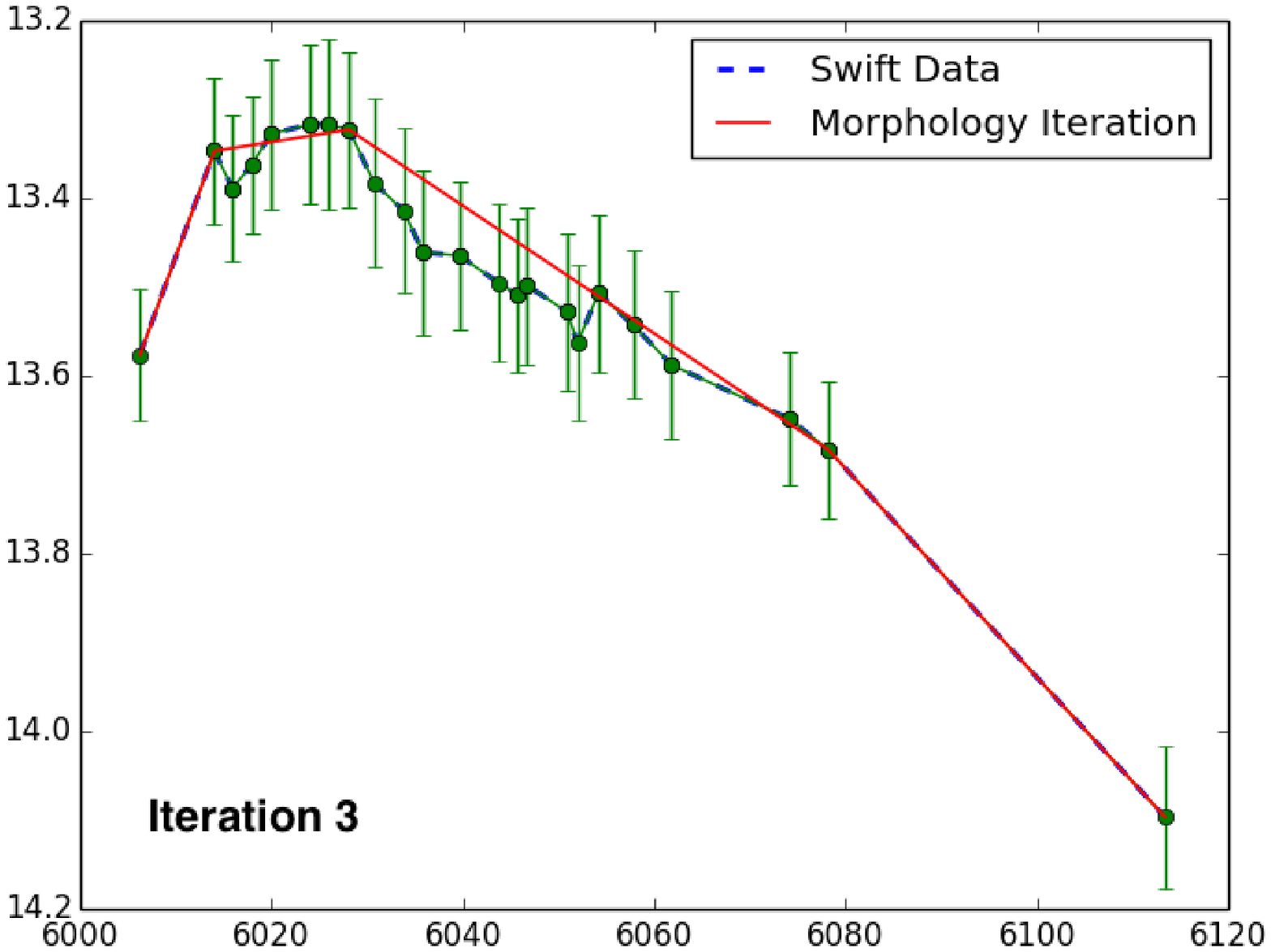}
\includegraphics[scale=0.45]{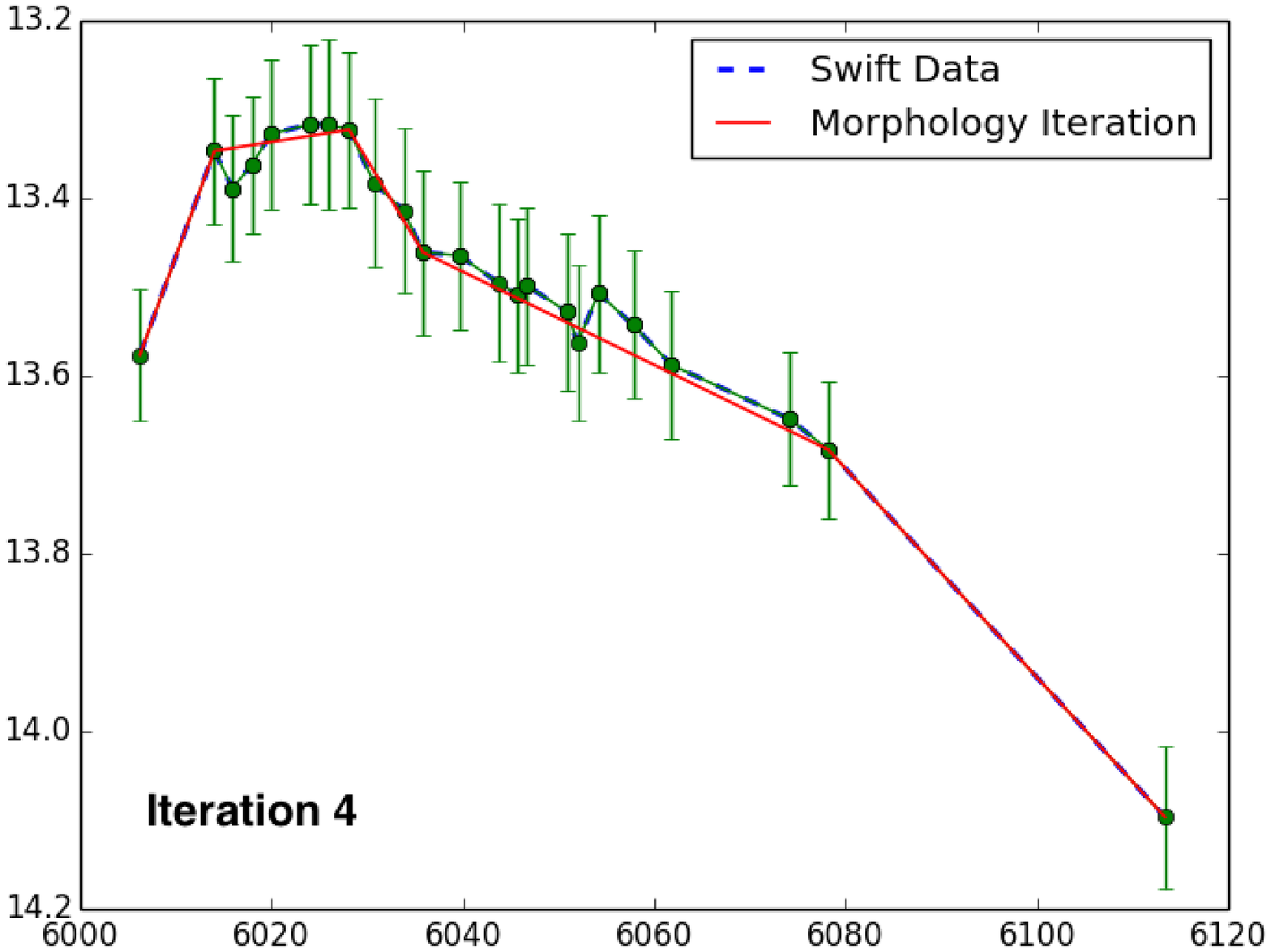}
\caption{Example for automating the characterization of light curve shapes. In this scheme, a line segment connecting the first and last data points is iteratively divided at points where the data differ the most on a given line segment.  The iteration terminates when deviations from the fit are within the error of the Swift data. The general shape of the curve is thus constrained robustly in the case of modest light curve noise. }
\label{lanl}
\end{figure*}

  Theoretically, the models in the database are not intended to match any one event, but be representative of a family of SN types. Thus, by shifting the model to match the data in time, we are learning more about how the models need altering for a faster or slower explosion evolution. 
The power in SNAP as an analysis tool is the ability to see not only what model-observation pairs work, but even more importantly which ones do not. These types of correlations between models and observations give us rapid feedback on what we are understanding and what physics we are missing in the models.  The SNAP database is, of course, a perpetual work in progress. The goal is to have a database containing models covering every type of CCSNe with reasonable parameter space coverage. Future versions will include additional extinction laws, account for error in distance and E(B-V), and return multiple interesting correlations based on selected criteria, including what sections of the light curve to focus on comparing. 
We are exploring better fits to observations in addition to the Gaussian process to describe light curves with largely spaced data, few data points, or have multiple slopes.  Figure \ref{lanl} shows an example of fitting the characteristics of an observed light curve which can then be compared to models.  Such schemes, which are currently being explored in SNAP, will allow one to quickly fit slopes for rise times, thermal decay and Ni-56 tails.

We also note that at present time we are not doing a Bayesian analysis.  One problem with the Bayesian analysis is that we do not really have any priors. For example, we can say that the mass of the progenitor is more than 8 \Msun, but what is the real mass range, energy range, radius range, etc. for Type IIPs? Type IIns? Type Ib/cs? Etc.  With SNAP we can determine what those priors are if there are models spanning parameter space from various sources and multiple sources of data in the system.  Just fitting one SN at a time is not telling us what is common in the entire family of SN types.

\section{The Web Application}
SNAP is accessed through a web application hosted at SwRI\footnote{The form to request access to the full SNAP system is available at snap.space.swri.edu.  Access is granted via a VPN account}.    From the web, users can download observations or model light curves and view correlations.  Current upload capability can be done through contact with the PI (A. Bayless), but will be available via the web application the next version of SNAP.   Correlations are a one-to-many relation, but certain filters on the correlation can be used. For example, a model-to-observation correlation may only include observations of Type IIP SNe and not all types.  An observation-to-model correlation may only include models within a certain progenitor mass range, for instance. In SNAP 1.0, users that wish to run correlations on a model set need to contact the PI until the full implementation of the web site is available.  Future versions of SNAP will allow users to upload their own observations or models and run their own correlations, with the option of saving them to the database, bypassing the current interface with the PI.   Permanent correlation additions will be cross-linked to the models and the observations and appear on the respective model and observation pages with the entry. For security, users will have editing ability on their data or models only, but will be able to view resulting model light curves (not source code) from other users, all telescope observations, which are public domain, and all correlations that have been run.  

\begin{figure*}
\center 
 \includegraphics[scale=0.45]{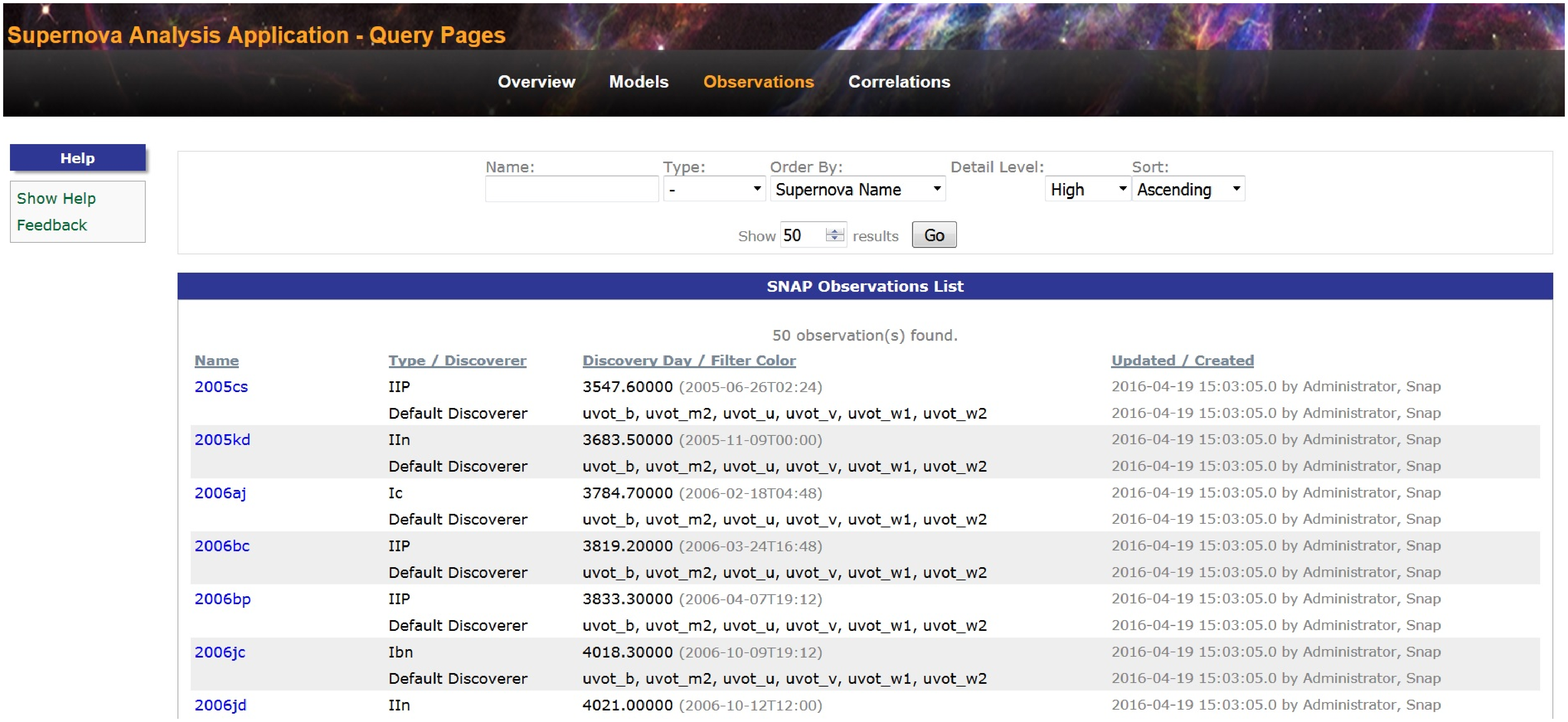}
\caption{Screen capture of the Observation Query page.  Users can search for SN names, by year, or type.  Each SN is a hyperlink that gives more detailed information, including distance, extinction, dates, coordinates, etc. and light curve information. The hyperlinks also cross reference any correlations that used that SN.}  
\label{obs}
\end{figure*}

\begin{figure*}
\center 
 \includegraphics[scale=0.75]{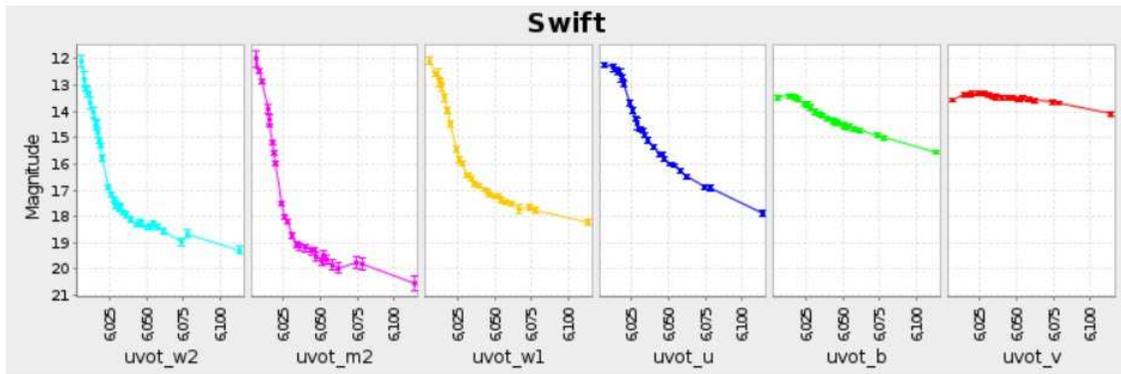}
\caption{Screen capture of the observation page for SN 2012aw. The page shows the available information and light curves, which can be downloaded as a text file by clicking on the light curve plot.}  
\label{obslc}
\end{figure*}

\begin{figure*}
\center 
 \includegraphics[scale=0.45]{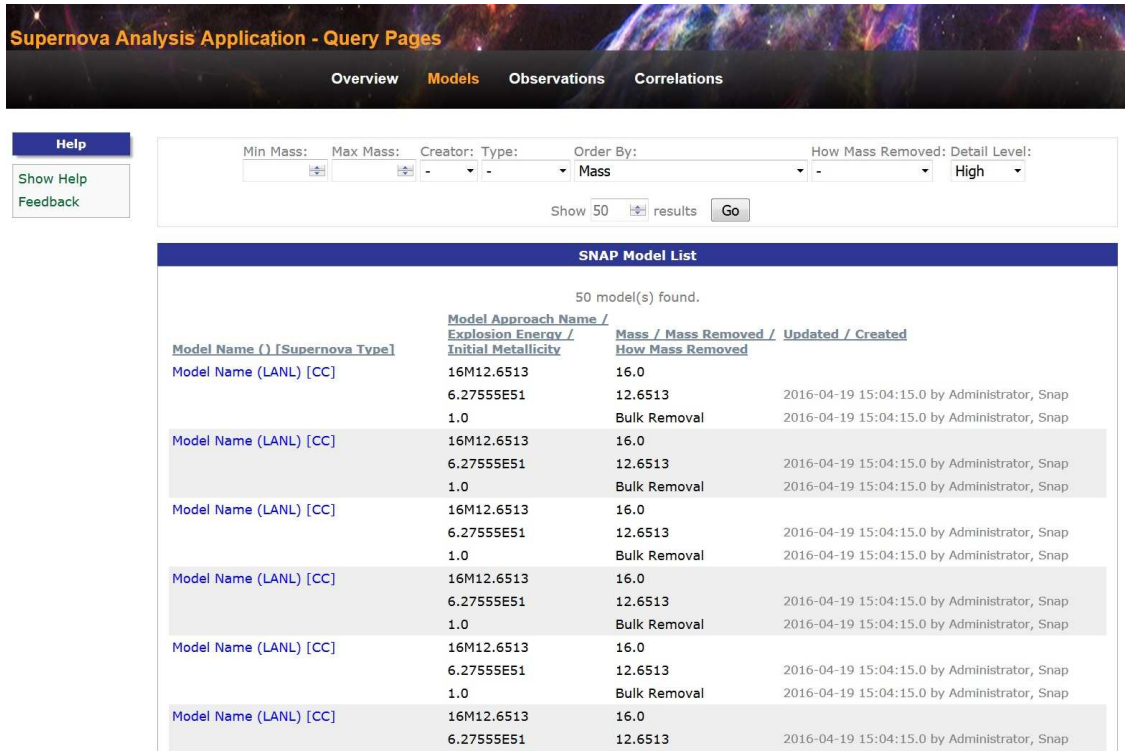}
\caption{Screen capture of the Model Query page.  Users can search for models by progenitor mass range, type, research group (creator of the model).  Each model is a hyperlink that gives more detailed information about the model parameters and light curves. The hyperlinks also cross reference any correlations that used that model.}  
\label{mod}
\end{figure*}

\begin{figure*}
\center 
 \includegraphics[scale=0.45]{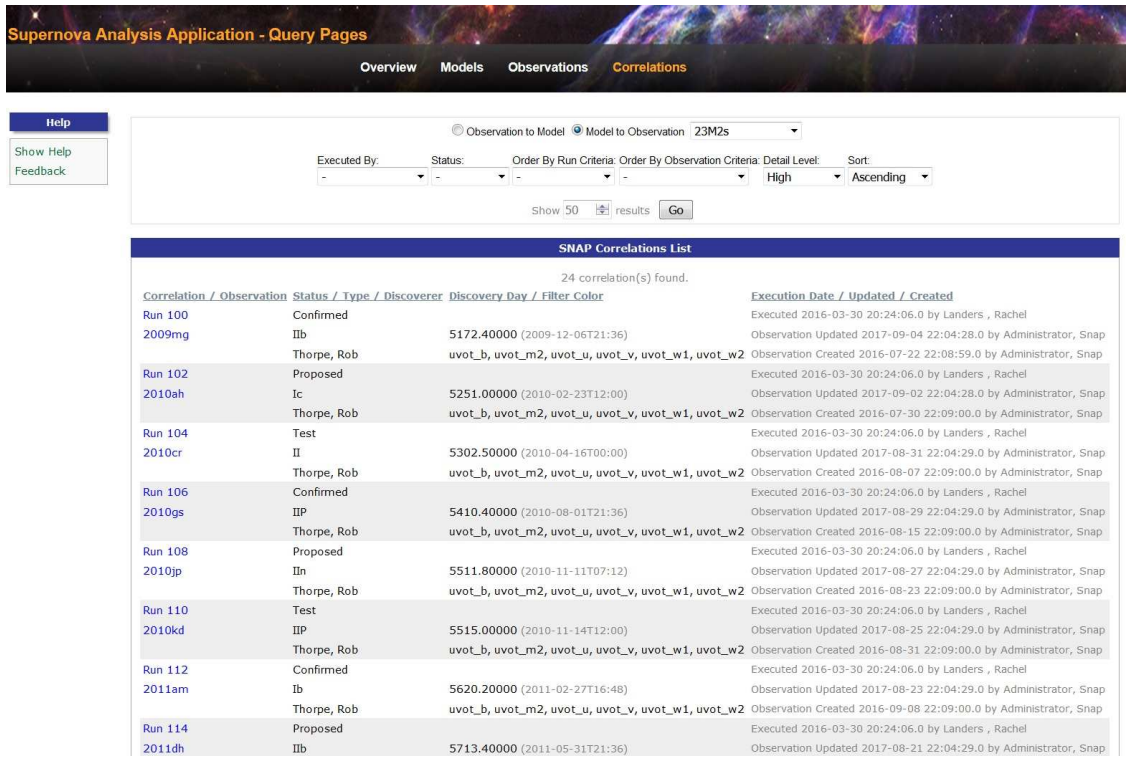}
\caption{Screen capture of the Model-to-Observation (M2O) Query page. Users can search for M2O correlations by creator of the correlation or status and ordered by several criteria options (date, SN name, status, etc.).  Each correlation is a hyperlink that gives cross linked information about the model's parameters and the SN observation that were correlated and an estimation on the goodness-of-fit, currently the delta slope described in the text.}  
\label{mod2obs}
\end{figure*}

\begin{figure*}
\center 
 \includegraphics[scale=0.45]{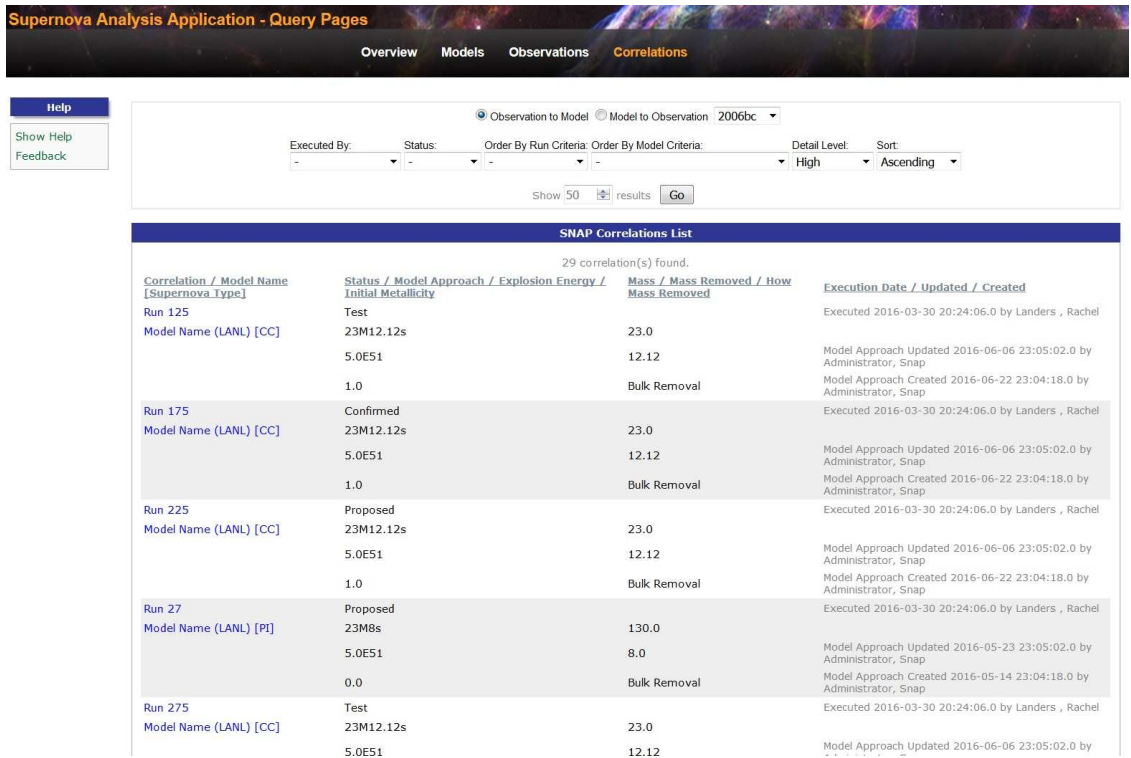}
\caption{Screen capture of the Observation-to-Model (O2M) Query page. Users can search for O2M correlations by creator of the correlation or status and ordered by several criteria options (creation date, SN type, etc.).  Each correlation is a hyperlink that gives cross linked information about the model's parameters and the SN observation that were correlated and an estimation on the goodness-of-fit, currently the delta slope described in the text.}  
\label{obs2mod}
\end{figure*}

The observational page has the options to search for a particular SN (with wild card options, e.g. 2012* returns all SN from 2012) or filter by a particular SN type.  Figure \ref{obs} shows a screen capture from the observation query page.  The high level returns the discovery date, MJD and calendar day, and the available filter band passes that have stored observations. Figure \ref{obslc} shows the entry for SN 2012aw including the plotted light curves.  clicking on the light curve plot allows users to download a text file of hte light curves giving JD, magnitude, and error in magnitude if available. The model page has the options to search for a range of progenitor star masses, the creator/research group, the SN type, and how mass was removed for Type IIb/Ib/Ic. Figure \ref{mod} shows a screen capture from the model query page. The correlation page allows users to compare and observations against a set of models or a model against a set of observations.  Figure \ref{mod2obs} shows a screen shot from the model-to-observation correlation query page. On this page there are options to search by correlation creator, status, SN type, and when the correlation was made.  The same options exist on the observation-to-model page (Figure \ref{obs2mod}), but also have added a filter by observational discovery day or shock breakout day (if known).  Any correlations that are saved between models and observations are cross linked to the model on model page and the observation on the observation page.  Thus, when searching for a particular model or observation, if there is a saved correlation, it will also be displayed in the search on those respective pages as well. 

\section{Conclusion} 
The era of large SN surveys will provide the opportunity to statistically study SNe in unprecedented numbers. Large surveys will also capture rare, exotic events which could contribute meaningfully to our understanding of the as yet enigmatic CCSN mechanism, which must be quickly identified among hundreds of other SNe for follow-up.  To study the broad spectrum of SN events in this flood of data, a means of rapidly and robustly characterizing transients against high-fidelity SN models is required. Additionally, the rapid improvements in computing are making more sophisticated models possible, which then require a means to validate these models.

In this paper, we have presented SNAP, a user-friendly database of SN observations and models with a novel fitting scheme which automates this process, accessed through a web browser.  The database has utility for studies of single events, allow for comparisons of  observations with an array of light curve models or vice versa, as well as large datasets of SN observations characteristic of upcoming surveys. We invite members of the community to contribute and use this database for their research needs. Indeed, the real power of SNAP is the cumulative ``network effects" of having users add models and data to the database through the web application -- as more models and data are added, more correlations will be possible giving more insight into SN behavior. 
SNAP will allow near real-time SN light curve fitting, eliminating in many instances the need to model events individually and assist observers in identifying the most interesting transients worthy of extended observations.

\acknowledgments
The authors would like to thank Daniel van Rossum for his work on the SuperNu code.  This work was supported by the Southwest Research Institute Internal Research program (Project Number R8333 \& R8498) and is supported by the NASA Astrophysical Data Analysis Program (NNH15ZDA001N-ADAP).  
\newpage


\newpage

\end{document}